\documentclass[12]{article}

\usepackage[dvipsnames]{xcolor}
\usepackage{color}
\usepackage{fancyvrb}
\usepackage{amsmath}
\usepackage{amssymb}
\usepackage{listings}
\usepackage{verbatim}
\usepackage{graphicx}
\usepackage{amsmath}
\usepackage{natbib}
\usepackage{amsmath,amssymb,amsfonts,latexsym,mathrsfs}
\usepackage{amsthm}
\usepackage{fullpage}

\theoremstyle{plain}
\newtheorem{thm}{Theorem}[section]
\newtheorem{lem}[thm]{Lemma}

\theoremstyle{definition}

\theoremstyle{remark}

\begin{document}

\def\P{{\bf P}}
\def\E{{\bf E}}
\def\IR{{\bf R}}
\def\IN{{\bf N}}
\def\II{{\bf I}}
\def\one{{\bf 1}}
\def\F{{\cal F}}
\def\G{{\cal G}}
\def\X{{\cal X}}
\def\Y{{\cal Y}}
\def\K{{\cal K}}
\def\C{{\cal C}}
\def\C{{\cal B}}
\def\M{{\cal M}}
\def\dd{{\cal D}}

\newcommand{\beq}{\begin{equation}}
\newcommand{\eeq}{\end{equation}}
\newcommand{\beqn}{\begin{eqnarray}}
\newcommand{\eeqn}{\end{eqnarray}}
\def\cas{{\cal S}}
\newcommand{\RR}{{\mathbf R}}
\newcommand{\rv}{\mbox{Var}}
\newcommand{\rcov}{\mbox{Cov}}
\newcommand{\ru}{{\mbox{Uniform}}}

\def\supp{{\rm supp}}
\def\eqref#1{(\ref{#1})}
\def\proof{\bigskip \noindent \bf Proof. \rm }

\author
{Radu V. Craiu \\
{\it email: craiu@utstat.utoronto.ca} \\
Department of Statistical Sciences\\
 University of Toronto, Toronto, ON, M5S 3G3, Canada.
\and
Thierry Duchesne\\
 {\it email: thierry.duchesne@mat.ulaval.ca} \\
D\'epartement de math\'ematiques et de statistique\\
Universit\'e Laval,
Qu\'ebec, QC, G1V 0A6, Canada}

\title{A scalable and efficient
covariate selection criterion for mixed effects regression models with unknown random effects structure}

\maketitle

\begin{abstract}

We propose a new model selection criterion for mixed effects regression models that is computable when
the model is fitted with a two-step method, even when
the structure and the distribution of the random effects are unknown. The criterion is especially useful in the early stage of the model building process when one  needs to decide which  covariates  should be included in a mixed effects regression model  but has no knowledge of the random effect structure. This is particularly relevant in substantive fields where  variable selection is guided by information criteria rather than regularization. The calculation of the criterion requires only the evaluation of cluster-level log-likelihoods and  does not rely on heavy
numerical integration.  We provide theoretical and numerical arguments to justify the method and we illustrate its usefulness by analyzing data on a socio-economic study of young American Indians.
\end{abstract}



\section{Introduction}

Studies where a large number of observations are collected for each experimental unit, or cluster, are quite common.
For instance, in behavioural ecology animals that wear GPS collars are tracked and data for each individual are collected
every hour for months or years; in marketing studies banks record every credit card transaction made by a client; in some
epidemiological studies data are collected on physicians who each treat  a large number of patients; in criminology,
data are recorded at every contact of a repeat  offender with the justice system. In the social study that we use to illustrate
our method, a large number of students are surveyed in a number of secondary high schools. In many instances where such data are collected,
analysts will account for the dependence within each cluster by  fitting a mixed effects regression model.  In the  construction of the latter an important and early step concerns selecting the covariates that are included in the model.

The importance of variable  selection has  been recognized in  statistics  and there is a vast body of work devoted to developing criteria for this problem \citep[e.g., see the book of][] {Burnham-Anderson02}.
Traditionally, the Akaike information criterion (AIC) introduced in the foundational work of \cite{akaike70}  along with its small sample corrections \citep{hurw-tsai,cava}, and the Bayesian Information Criterion (BIC), introduced by
\cite{Schwarz78}, have been among the first methods used to select the covariates in regression models with fixed effects. All these are special cases of the  Generalized Information Criterion (GIC)  \citep{nishii,shibata,rao-wu} where the  aim is to find the model $M$ that minimizes \beq -{\cal{L}}(M) + \lambda |M|,\label{gic}\eeq where ${\cal{L}}(M)$ is a measure of fit and $\lambda|M|$ is the  penalty incurred by a model with size $|M|$. The GIC proposed by \cite{rao-wu}
 is a strongly consistent variable selection criterion with a flexible penalty function.


The introduction of mixed effects models  required new strategies for selecting both the fixed and the random effects. In this context, whether the  inferential focus is on marginal or conditional model parameters becomes relevant as these two scenarios  require separate treatments. While in the former case one could use the traditional  criteria to select the covariates in the model, the latter considers the choice of covariates conditional on random effects.  In \cite{vaid:blan:05} the authors proposed the conditional AIC (cAIC) for situations in which the inferential focus is on cluster-specific parameters. Subsequently, the cAIC for linear mixed models has been further expanded by \cite{liang-wu}, \cite{greven-kneib} and \cite{saefken} who account for the estimation of variance parameters and by \cite{donohue} who have  extended cAIC to generalized linear mixed models (GLMM) and survival models with random effects. \cite{YuEtAl13} have proposed a further
adjustment for cAIC in GLMM when the variance components must be estimated.
An alternative BIC suitable for mixed effects models has been proposed by \cite{delattre}. In a departure from classical approaches, \cite{jiji08} propose a  method in which incorrect models are fenced off and the best model is selected from the remaining ones.  An excellent review of the methods briefly discussed here and  others can be found in
\cite{ms-lmm}.



Our current contribution for a new  criterion is motivated  by  GLMM applications in ecology and social sciences where model selection is traditionally based on information criteria and not on regularization methods. Moreover, in these fields
little is known about the structure of the random effects {\it a priori} and  numerical approximations  of the marginal likelihood may be challenging due to model and data size \citep{cdfb,molen2011}.   The new criterion is intended as a first covariate filter in the early stage of the analysis. Given this aim,  it is important that the proposed criterion is computable without the need to specify the random effect structure and that one does not incur a heavy computational cost in order to  calculate  its value for the submodels under consideration.
After this initial stage, other numerically demanding methods
such as the marginal AIC, marginal BIC or cAIC  can be exploited to search in the smaller model space.

The key strategy in the derivation of this new criterion is to avoid  computation of the marginal log-likelihood function or
global maximum likelihood estimators, which requires the specification of a random effect structure and the use of numerical integration.
The criterion developed here is suitable for   ``partitioned data'' methods (sometimes referred to as ``divide-and-conquer'' approaches)
that have been proposed to fit mixed effects models when the data are large or have a complex structure. Such methods include the
two-stage approach of \cite{Korn:1979vn} and \cite{Stiratelli:1984uq}, the CREML method
of \cite{Chervoneva:2006lr}, the two-step method of \cite{cdfb} or the pseudo-likelihood approach of \cite{molen2011}.
All these methods have in common that they fit separate
simple models to each element of a partition of the data and then suitably unify the analyses for these simple models to produce inference for the global mixed effects model. Since these approaches do not resort to numerical integration, they make it easier to avoid computational issues and they can even remain asymptotically
efficient in some specific cases.

In this paper we focus on deriving a criterion for filtering the potential covariates for use in a standard GLMM as described, for instance, in  Chapter 3 of \cite{jiji}.
The proposed criterion, called meanAIC, is  easy to compute  and it does not require the specification of the random effects structure.
We give a theoretical development of meanAIC along with heuristic arguments that justify it.
Our simulation study shows that the proposed criterion exhibits good finite sample performance.

The remainder of the paper is organized as follows. Section \ref{sec:datglmm} presents the data and model. The new criterion is developed and justified
in Section \ref{sec:crit}. The simulation study is presented in Section \ref{sec:sim} and a real data illustration forms Section \ref{sec:eg}. The paper concludes with a discussion and ideas for future work.

\section{Data and model}\label{sec:datglmm}

\subsection{Population and data}

We consider a population  of independent clusters, each containing  a number of individual observations of the form $(Y,x_{1},\ldots,x_{p})$ with $Y$ being a response variable and $x_1,\ldots,x_p$ potential
explanatory variables. We assume that the distribution of $Y$ given $x_1,\ldots,x_p$ is given by a generalized linear model whose
regression coefficients may vary from cluster to cluster.
The statistical model described below will assume that all the responses in the same cluster share some commonality that  makes them dependent.  We  assume that there are $K$ clusters and $n_i$ data points in each cluster, $1\le i \le K$.

\subsection{Generalized linear mixed model (GLMM)}

 Let $Y_i=(Y_{i1},\ldots,Y_{in_i})^\top$ be the response vector for cluster $i$ and
$X_i=(x_{0i}$, ${x}_{i1}$, $\ldots$, ${x}_{ir})$
be the corresponding covariate matrix, with ${x}_{ik}=(x_{i1k}$, $\ldots$, $x_{in_ik})^\top$, $k=0,\ldots,r$ and
$x_{0i}$ an $n_i$-vector with all entries equal to 1. Throughout the paper the value of the $k$th covariate   for the $j$th individual in the $i$th cluster will be denoted $x_{ijk}$. The context will clarify whether  $x_{ij}$ refers to the $j$th row of $X_i$ (of length $r$) or  $x_{ik}$ refers to the  $k$th column of $X_i$ (of length $n_i$).

The dependence among observations in cluster $i$ will be captured using the random vector $b_i$, where $\{ b_i \in \RR^q\; : \; i=1,\ldots,K\}$ are  assumed to be iid  with
cumulative distribution function (cdf) $H$ and probability density function (pdf) $h$. Throughout the paper
we suppose that $q\le r$. Let ${\cal{J}}$ be a subset of size $s$ of $\{0,\ldots,q\}$, $Z_i=\{{x}_{ik},k\in{\cal{J}}\}$ and
$\beta=(\beta_0,\ldots,\beta_r)^\top$. Under our population assumption and sampling scheme: i) $(Y_i,X_i)$, $i=1,\ldots,K$, are independent and ii) for all $1\le i \le K$   $\{Y_{ij} \; : 1\le j \le n_i\}$ are independent given $X_i$ and $b_i$, iii) the distribution of $Y_{ij}|b_i,X_i$ belongs to the exponential family with pdf $f_{ij}$ and
\begin{equation}\label{eq:muglmm}
\mu_{ij}=E[Y_{ij}|b_i,X_i] = g^{-1}(\beta^\top {x}_{ij}+b_i^\top z_{ij}),
\end{equation}
where $x_{ij}^\top$ and $z_{ij}^\top$  denote the $j$-th row of $X_i$ and  $Z_i$, respectively, and  $g$ is a known link function.
This is the usual GLMM with random regression coefficients \citep[Ch. 3 in][]{jiji}. Let $\tilde{x}_{ij}=x_{ij}+\tilde{z}_{ij}$,
where $\tilde{z}_{ijk}={z}_{ijk}$ if $k\in{\cal{J}}$ and 0 otherwise. Similarly, let $\tilde{\beta}_i=\beta_i+\tilde{b}_i$
with $\tilde{b}_{ik}=b_{ik}$ if $k\in{\cal{J}}$ and 0 otherwise. Then (\ref{eq:muglmm}) can be written as
\begin{equation}\label{eq:muglmm2}
\mu_{ij}=E[Y_{ij}|b_i,X_i] = g^{-1}(\tilde{\beta}_i^\top \tilde{x}_{ij})
\end{equation}
and the average conditional log-likelihood contribution in cluster $i$ can be written as
\begin{equation}\label{eq:elli}
\bar{\ell}_{n_i}(\tilde{\beta}_i)=n_i^{-1}\sum_{j=1}^{n_i}\log f_{ij}(Y_{ij};\tilde{\beta}_i).
\end{equation}

When one wishes to keep all $r$ covariates in the model, $H$ is the multivariate normal with mean 0 and variance matrix $D$
and the subset ${\cal{J}}$ and the structure of the matrix $D$ are known, then
inference methods for $\beta$ and $D$, as well as predictions for $b_i$, are  widely available in standard
software. They are typically based on standard maximum likelihood, residual maximum likelihood, penalized quasi-likelihood or
Bayesian methods. However, in many practical situations all $r$ covariates are not required and the random effect structure (subset ${\cal{J}}$)
and distribution ($H$) are not known. In the following section we will derive a criterion that can guide  the selection of covariates  without having to specify either ${\cal{J}}$ or $H$ { and that can be computed when inference about $\beta$ is performed via a two-step
method.}

\section{The meanAIC criterion}\label{sec:crit}

When fitting a mixed effects model,  \cite{vaid:blan:05} and \cite{greven-kneib} argue that the selection of fixed effects and other marginal population parameters can be performed using the marginal Akaike information
criterion (henceforth denoted mAIC). However, mAIC  is based on the maximized marginal log-likelihood, whose computation
involves marg\-inal\-izing out the random effects via numerical integration.
A  challenge is that the latter may become numerically cumbersome when the dimension $s$ of the random effects is large or when the data are massive. A  {potentially} more serious problem is that marginal likelihood calculation requires the specification of the random
effect structure ${\cal{J}}$ and distribution $H$. {As we shall see in the real data illustration, the covariates to be selected
by mAIC may vary according to assumptions made about the random effect structure.}

When considering a partitioned data approach, one realizes that the GLMM specification yields ordinary independent GLMs
in each cluster. Many two-stage estimation approaches \citep[e.g.,][]{Stiratelli:1984uq,molen02,Chervoneva:2006lr,cdfb,molen2011} rely on
(some of) the following cluster-specific information that is usually easier to obtain than their full data counterparts:
\begin{itemize}
\item $n_i$, the number of observations in cluster $i$;
\item $\hat{\beta}_i=\arg \max_\beta \bar \ell_{n_i}(\beta)$, the MLE of $\beta$ in cluster $i$;
\item $\bar{\ell}_{n_i}(\hat{\beta}_i)$, the maximized average log-likelihood in cluster $i$;
\item $H_i=\left.\frac{\partial^2}{\partial\beta\partial\beta^\top}\bar{\ell}_{n_i}(\beta)\right|_{\beta=\hat{\beta}_i}$, the Hessian
of the average log-likelihood evaluated at $\hat \beta_i$.
\end{itemize}
It is clear that a criterion that would be based only on these four elements from each cluster should be easy to compute in practice. When the number of clusters is large, computations can be easily parallelized with one CPU  fitting an ordinary GLM
to its assigned cluster and returning the required cluster-level objects.

\subsection{Derivation of meanAIC}

In our derivations we work under the assumption that all cluster sizes,  $n_i$, are large.
This condition is  generally satisfied  when two-step inference methods are used \citep{molen2011}. The first step
 amounts to fitting an ordinary GLM to the data from each cluster. Under our model assumptions, the
GLM in cluster $i$ has regression parameter $\tilde{\beta}_i=\beta+\tilde{b}_i$ and because the $\tilde{b}_i$ are iid from a continuous distribution,
all $\tilde{\beta}_i$ have the same zero elements with probability 1. Let us assume that the true data generating model
is the GLMM considered evaluated at $\beta=\beta^0$ and let $\tilde{\beta}_i^0=\beta^0+\tilde{b}_i$, i.e.
 the true data generating model is among the models considered.  Let
\beq
\Delta_i(\beta|\tilde{b}_i)=
-\int \log\left\{{f_{ij}(Y_{ij};\tilde{\beta}_i)}\right\}f_{ij}(Y_{ij};\tilde{\beta}_i^0)dY_{ij}
\label{KLoverall}
\eeq
be the conditional cluster-level Kullback-Leibler divergence
for a GLM with parameter $\beta$ in cluster $i$ and put $\Delta(\beta|\tilde{b}_1,\ldots,\tilde{b}_K)=K^{-1}\sum_i\Delta_i(\beta|\tilde{b}_i)$.
Conditional on the random effects, we consider the cluster-specific empirical  versions of the above divergences,  respectively given by
\begin{align*}
\hat{\Delta}_i(\beta|\tilde{b}_i)&=-n_i^{-1}\sum_j\log\left\{{f_{ij}(Y_{ij};\tilde{\beta}_i)}\right\}=-\bar{\ell}_i(\beta+\tilde{b}_i)\\
\hat{\Delta}(\beta|\tilde{b}_1,\ldots,\tilde{b}_K)&=K^{-1}\sum_i\hat{\Delta}_i(\beta|\tilde{b}_i)=-K^{-1}\sum_i\bar{\ell}_i(\beta+\tilde{b}_i).
\end{align*}

The following lemma, whose simple proof is sketched in the appendix, establishes some useful properties of
$\Delta_i(\beta|\tilde{b}_1i)$.

\begin{lem}\label{lem:1}
Under the GLMM assumptions made in Section 2, $\Delta_i(\beta|\tilde{b}_i)$ is a smooth function of $\beta$ uniformly in $\tilde{b}_i$.
Furthermore, $\Delta_i(\beta|\tilde{b}_i)\ge\Delta_i(\beta^0|\tilde{b}_i)$ for all $\beta$ for any finite value of $\tilde{b}_i$.
\end{lem}

The most important corollary to Lemma \ref{lem:1} is that $\Delta(\beta|\tilde{b}_1,\ldots,\tilde{b}_K)$ is minimized
at $\beta=\beta^0$ regardless of the values of $\tilde{b}_1,\ldots,\tilde{b}_K$. It can therefore serve as a covariate selection
criterion if an asymptotically unbiased estimator of the expectation of its empirical version is available. To this end, Remark 1
on p. 242 in the Appendix of \cite{lin-zuc} translates into the following lemma.
\begin{lem}\label{lem:2}
\begin{align}
\label{eq:lem21}\hat{E}[\hat{\Delta}_i\{\hat{\beta}(\tilde{b}_i)|\tilde{b}_i\}|\tilde{b}_i]&=
\hat{\Delta}_i\{\hat{\beta}(\tilde{b}_i)|\tilde{b}_i\}+\frac{r+1}{n_i}\\
\label{eq:lem22}\widehat{Var}[\hat{\Delta}_i\{\hat{\beta}(\tilde{b}_i)|\tilde{b}_i\}|\tilde{b}_i]&=\frac{(r+1)/2}{n_i^2}.
\end{align}
\end{lem}
In equations (\ref{eq:lem21}) and (\ref{eq:lem22}) $\hat{E}$ and $\widehat{Var}$ respectively denote consistent estimators of the mean and variance of the empirical divergences
\citep[p. 242]{lin-zuc}.

Because the AIC in cluster $i$ is given by $2n_i\{\hat{\Delta}_i\{\hat{\beta}(\tilde{b}_i)|\tilde{b}_i\}+{(r+1)}/{n_i}\}$, equation (\ref{eq:lem21})
establishes that the best model should have smallest AIC in all clusters. Thus, when all the $n_i$ are large we can minimize a  weighted
average of the cluster-level AICs to  identify the  model that minimizes \eqref{KLoverall}. According to equation (\ref{eq:lem22}), the
weighted average of the form $\sum_i w_i 2n_i\{\hat{\Delta}_i\{\hat{\beta}(\tilde{b}_i)|\tilde{b}_i\}+{(r+1)}/{n_i}\}$ with $0\le w_i\le 1$
and $\sum_iw_i=1$ with the smallest variance has $w_i=1/K$, suggesting the following model selection criterion:

\beq
\mbox{meanAIC}={1\over K} \sum_{i=1}^K \text{AIC}_i,
\label{meanAIC}
\eeq
where $\mbox{AIC}_i = -2 \log f_i(y_i ; \hat{\beta}_i) + 2(r+1)$.
Thus, meanAIC is
simply the average of all $K$ AICs obtained when fitting an ordinary GLM separately to each of the $K$ clusters.

{The derivation of the meanAIC is based on the Kullback-Leibler distance between the cluster-specific densities defined by the stage 1 estimation procedure. Although it deviates from the canonical elicitations of AIC-type criteria, we believe it is the only way we can bypass the specification of random effects distribution or structure, e.g.  identification of covariates with  random effects. }

\section{Simulation study}\label{sec:sim}

The meanAIC criterion was derived in the previous section as a potentially useful
covariate selection tool when the cluster sizes $n_i$ tend to infinity. The present section reports
the results of a simulation study whose primary objective is to assess the performance of meanAIC
as a covariate screening tool for finite values of $n_i$. A secondary objective is
to compare its efficiency and robustness to that of mAIC that is computed assuming a GLMM with a random intercept. The latter choice is in line with our aim of establishing model selection criteria without having to spell out the random effects structure; under this constraint, the only specification that is common to all mixed effects submodels is the one with only a random intercept.

\subsection{Study design}


We generated samples with $K=20$ independent clusters. Our primary objective suggests that we consider larger values of $n_i$.
We ran simulations where $n_i$ was fixed at 80 or 320 observations in all clusters; additional simulations
where $n_i$ varied from cluster to cluster within the same dataset yielded results similar to those reported below
and are summarized in Appendix A.
Our simulation design is similar to other designs where model selection criteria are investigated
\citep{YuEtAl13}.
We simulate  two covariates,
$x_{ij1}$ iid Bernoulli$(0.5)$ and $x_{ij2}$ iid uniform$(0,1)$. The responses $Y_{ij}$ are  generated from a Poisson
GLMM with log link and depend on $x_{ij1}$, a random intercept $b_{0i}$ and a random coefficient
$b_{1i}$ through the following log conditional mean:
\begin{equation}\label{eq:simmodel}
\log E[Y_{ij};b_{0i},b_{1i}]=0.3+b_{0i}+(\beta_1+b_{1i}) x_{ij1},
\end{equation}
for $1\le i \le K$ and $1\le j \le n_{i}$.

Our simulations consider two values for the fixed effects  $\beta_1\in\{0.2,0.4\}$ and
 assumed the two random effects to be independently drawn from the same  zero-mean normal distribution, but with possibly different variances,
more precisely $b_{ui}$ has variance $\sigma_u^2$ which can take values in the set $\{0.005$, $0.15$, $0.3$, $0.8$, $1.5\}$ for all $u\in\{0,1\}$ .  We also simulate random effects from gamma and t distributions in order to investigate robustness of the criteria against asymmetric and fat-tailed  distributions, respectively. These results
are quite similar to those reported in Table \ref{tab:n80beta02normal} and are summarized in Appendix A.

For each sample generated, all four submodels were considered: i) the null model; ii) the true model with $x_1$ only; iii) the model with $x_2$ only, and iv) the model with both $x_1$ and $x_2$.
For each sample the following model selection criteria were computed:
the proposed meanAIC and
the mAIC obtained by fitting a Poisson GLMM with a random intercept to the entire sample.

The meanAIC is calculated using the  output of the {\tt glm} function from R
applied to each cluster separately. Maximum likelihood fitting of the random intercept GLMM was implemented
with the function {\tt glmer} from the R package {\tt lme4}. The calculation for meanAIC was approximately three times faster than the mAIC, e.g., computation for one sample required 0.05 second of CPU time for meanAIC and 0.17 second of
CPU time for the mAIC on a Lenovo X230 tablet PC with Intel Core i7-3520M CPU at 2.90 GHz, 8 GB of RAM and running on the 64 bit
version of Windows 7.

\subsection{Results}

Each simulation scenario was replicated 500 times and the proportion of correct covariate selection decisions are
reported in  Table \ref{tab:n80beta02normal}. The simulations show that mAIC dominates meanAIC only  when cluster-sizes are small and the random coefficient has a small variance. This is not surprising given that the mAIC criterion is computed assuming that only the intercept is random. Therefore, when averaging over the distribution of the random intercepts, mAIC pools all the cluster data and benefits from the resulting larger sample size, unlike meanAIC which relies on cluster-level data sizes that are not large enough  to enable it to detect the small fixed effects. As soon as cluster-specific coefficients are not all small, e.g. the random coefficients have moderate variance, the accuracy of meanAIC dramatically increases. For example, when  $n_i=80$ and the random effect variance $\sigma_1^2$ increases from $0.005$ to $0.15$  we see that the percentage of correct decisions for meanAIC increases from about 12\% to about 90\% for different values of $\sigma_0^2$. This trend is not replicated by mAIC that does not seem to benefit from larger values of $\sigma_1^2$. When $n_i=320$ both meanAIC and mAIC have a good performance, with meanAIC dominating mAIC for all values of $\sigma_0^2$ and $\sigma_1^2$ considered.

\begin{table}
\caption{Proportion of the 500 replications where each criterion picked the true model.
Regression coefficient $\beta=0.2$.
Random effects follow a normal with mean 0
and variance $\sigma^2_u$, $u=0,1$.}\label{tab:n80beta02normal}
\smallskip
\begin{center}
\begin{tabular}{ccllrr}\hline
 & & \multicolumn{2}{c}{$n_i=80~\forall i$} & \multicolumn{2}{c}{$n_i=320~\forall i$} \\
$\sigma_0^2$ & $\sigma_1^2$ & meanAIC & mAIC & meanAIC & mAIC   \\  \hline
0.005 & 0.005 & 0.140 & 0.790 & 0.878 & 0.828  \\
0.005 & 0.15 & 0.884 & 0.740 & 0.994 & 0.828 \\
0.005 & 0.3 & 0.980 & 0.778  & 0.994 & 0.830  \\
0.005 & 0.8 & 0.992 & 0.816  & 0.994 & 0.774 \\
0.005 & 1.5 & 0.994 & 0.772 & 0.996 & 0.794  \\
0.15 & 0.005 & 0.166 & 0.792 & 0.876 & 0.846  \\
0.15 & 0.15 & 0.882 & 0.774 & 0.996 & 0.818  \\
0.15 & 0.3 & 0.988 & 0.786 & 0.998 & 0.838  \\
0.15 & 0.8 & 0.992 & 0.788 & 0.996 & 0.794  \\
0.15 & 1.5 & 0.996 & 0.746 & 0.992 & 0.762  \\
0.3 & 0.005 & 0.116 & 0.802 & 0.932 & 0.844  \\
0.3 & 0.15 & 0.894 & 0.732 & 0.998 & 0.836  \\
0.3 & 0.3 & 0.988 & 0.788 & 0.990 & 0.822  \\
0.3 & 0.8 & 0.998 & 0.754  & 0.990 & 0.798 \\
0.3 & 1.5 & 0.998 & 0.730 & 0.992 & 0.738  \\ \hline
\end{tabular}
\end{center}
\end{table}

To see if the performance of meanAIC with $\sigma_1^2=0.005$ and $n_i=80$ improves as the fixed effect size $\beta$ gets larger, we replicate the simulation scenarios involving these values of $n_i$ and $\sigma_1^2$, but with a larger  $\beta=0.4$.
The results are summarized in Table \ref{tab:n80beta04normal}
 and show that under these settings meanAIC outperforms mAIC.

\begin{table}
\caption{Proportion of the 500 replications where each criterion picked the true model.
Regression coefficient $\beta=0.4$.
Random effects follow a normal with mean 0
and variance $\sigma^2_u$, $u=0,1$.}\label{tab:n80beta04normal}
\smallskip
\begin{center}
\begin{tabular}{ccll}\hline
 & & \multicolumn{2}{c}{$n_i=80~\forall i$}  \\
$\sigma_0^2$ & $\sigma_1^2$ & meanAIC & mAIC   \\  \hline
0.005 & 0.005 & 0.866 & 0.806  \\
0.15 & 0.005 & 0.910 & 0.856  \\
0.3 & 0.005 & 0.928 & 0.854  \\ \hline
\end{tabular}
\end{center}
\end{table}

\section{Real Data Application}\label{sec:eg}

We illustrate the ability of meanAIC to identify those covariates that can have strong cluster-specific effects, but small marginal effects. These are typically  covariates corresponding to important random effects variances that have modest fixed regression coefficients. The simulation studies performed in the previous section suggest that under this  scenario mAIC and meanAIC are likely to select different models.   The data come from  the study of  \cite{beauv} on alcohol use among young American Indians in U.S. schools.
The individual observations are responses by students to a survey, and
the clusters are the schools the students belong to. The sample we analyze consists of data from the  25 largest clusters with an average size of
170 students. The response variable  is the number
of times  a student had more than 5 alcoholic drinks in less than two hours during the last two weeks.

All models considered  will include four personal covariates (gender, age, whether the student likes school  and whether the student is proud of him/herself), two covariates related to friends influence (whether some of the friends have ever been suspended from school  and whether
friends ask the student to get drunk some or a lot).
The four candidate models differ due to presence/absence of two family-related covariates (family is a lot likely to stop the student from getting drunk, and  the family members argue a lot). A more detailed description of the model covariates and their corresponding values is presented in Appendix B.

The models are fit using  a Poisson GLMM with log link, with each  school as a cluster.
The meanAIC is obtained by fitting an ordinary Poisson GLM separately to each cluster data using the {\tt glm} function
in R.
The mAIC is computed by fitting the  marginal model using the   {\tt glmer} function from package {\tt lme4} in R \citep{lme4}.
Unlike meanAIC, to fit the GLMM needed for mAIC, one needs to specify a random effects structure. We  consider two such structures: i) a random intercept only
that will yield an mAIC value, denoted mAIC.RI, and ii)  a random intercept and a random coefficient for the ``family members argue a lot''
covariate that will yield an mAIC value, denoted mAIC.RC. A summary of these GLMM fits is provided in Appendix B. Clearly, the random coefficient for ``family members argue a lot'' has a large variance and a small marginal effect.  It is thus expected that meanAIC  may  identify more accurately than mAIC.RI the importance of this covariate  for the model.
The values of mAIC for the two GLMM and of meanAIC are reported in Table \ref{tab:critapplic}.
Simulations showed that meanAIC is better than mAIC.RI   at identifying the generating model covariates  that had a random coefficient. Based on these results and the magnitude of the random coefficient variance (this variance is highly significant according to the likelihood ratio test described in Section 6.3.2
of \cite{VerbMol2009})
 reported in Appendix B, we believe that ``family members argue a lot'' should be part of the model. All criteria agree that the family is ``likely to stop you from getting drunk" variable should be included in the model. It is worth pointing out that mAIC chooses different models depending on the random effects structure assumed, which can be confusing when there is no clear choice for the latter.

\begin{table}
\caption{Model selection criteria for all four submodels of interest for the alcohol consumption study.
mAIC.RI refers to the mAIC criterion obtained by fitting a GLMM with random intercept only, mAIC.RC denotes
the mAIC of the model with a random intercept and a random coefficient in front of the covariate ``family
members argue a lot'', while meanAIC denotes the meanAIC criterion.
The best value for each criterion appears in bold.
}\label{tab:critapplic}

\smallskip
\begin{center}
\begin{tabular}{cccc}\hline
Family covariates &  & & \\
in model &  mAIC.RI & mAIC.RC & meanAIC \\ \hline
Both & 7929.87 & {\bf 7898.47} & {\bf 298.47} \\
``Lot likely to stop you'' only & {\bf 7928.37} & 7928.37 & 302.26 \\
``Family members argue a lot'' only & 8166.20 & 8127.02 & 307.04 \\
None & 8166.06 & 8166.06 & 311.33 \\
\hline
\end{tabular}
\end{center}
\end{table}

\section{Discussion}\label{sec:disc}

In this paper we set out to develop a new variable  selection criterion for GLMM that does not require a specification of the random effects structure. Furthermore, we wanted a criterion computable even when a two-stage estimation procedure is used to fit the model, which usually occurs when the cluster sizes are large enough to make impractical  marginal likelihood inference.

We used an h-likelihood based  theoretical justification to develop the meanAIC criterion. The implicit assumption is that cluster sizes are large. Simulations were performed under a number of possible combinations of cluster size (small and large), random coefficients variance (small, moderate and large), effect size (small and moderate) and different random effects distributions (normal, shifted gamma and t). We compared the ability  of  meanAIC and of mAIC   to identify the true  covariate structure. The proposed meanAIC clearly outperformed mAIC for all settings except the case where cluster size, random coefficient variance and fixed effect size were all simultaneously small. The application of these criteria to real data analysis further emphasized the importance of variable selection without specifying the random effects structure.

Model selection based on comparing all possible  submodels is not practical when the number of potential covariates is large. In future work we would like to consider the interplay between meanAIC and regularization-based methods  \citep[e.g.,][]{ibra11, fanli12,lpj2013}, where information criteria are used to set the value of the tuning parameter in the penalty term.

\section*{Acknowledgments}

This work has been funded by  Natural Sciences and Engineering Research Council of Canada individual discovery  grants to each author.

\bibliographystyle{ims}
\bibliography{superref}

\appendix

\section{}
\subsection{Sketch of the proofs of lemmas \ref{lem:1} and \ref{lem:2}}
We have that $\Delta_i(\beta|\tilde{b}_i)=-\bar{\ell}_i(\beta+\tilde{b}_i)$, which is, up to the constant $-n_i$,
the log-likelihood of an ordinary GLM. The usual proof of the information inequality leads to the conclusion of
Lemma \ref{lem:1}:
\begin{align*}
\Delta_i(\beta|\tilde{b}_i)&=-\int \log\frac{f(y;\beta|\tilde{b}_i)}{f(y;\beta^0|\tilde{b}_i)}f(y;\beta^0|\tilde{b}_i)\;dy
\le -\log\int \frac{f(y;\beta|\tilde{b}_i)}{f(y;\beta^0|\tilde{b}_i)}f(y;\beta^0|\tilde{b}_i)\;dy
=0=\Delta_i(\beta^0|\tilde{b}_i).
\end{align*}

{Lemma \ref{lem:2} is simply the result of Remark 2 on p. 242 of \cite{lin-zuc} written using the notation of this paper,
which we can invoke when the GLM in each cluster meet standard GLM regularity conditions \citep[see, for instance, page 346 of][]{FK1985}}

\subsection{Additional simulation results}

\begin{table}
\caption{Proportion of the 500 replications where each criterion picked the true model.
Regression coefficient $\beta=0.2$.
Random effects follow a shifted gamma distribution with shape parameter 4, mean 0
and variance $\sigma^2_u$, $u=0,1$.}\label{tab:n80beta02gamma}
\smallskip
\begin{center}
\begin{tabular}{ccrlrc}\hline
 & & \multicolumn{2}{c}{$n_i=80~\forall i$} & \multicolumn{2}{c}{$n_i=320~\forall i$} \\
$\sigma_0^2$ & $\sigma_1^2$ & meanAIC & mAIC & meanAIC & mAIC   \\  \hline
0.005 & 0.005 & 0.072 & 0.780 & 0.856 & 0.856  \\
0.005 & 0.15 & 0.860 & 0.784  & 0.996 & 0.856 \\
0.005 & 0.3 & 0.962 & 0.778  & 0.988 & 0.798 \\
0.005 & 0.8 & 0.996 & 0.740  & 0.994 & 0.764\\
0.005 & 1.5 & 0.998 & 0.720  & 0.990 & 0.722\\
0.15 & 0.005 & 0.136 & 0.780  & 0.848 & 0.860 \\
0.15 & 0.15 & 0.862 & 0.758  & 0.994 & 0.836 \\
0.15 & 0.3 & 0.956 & 0.746  & 1.000 & 0.808 \\
0.15 & 0.8 & 0.994 & 0.694  & 0.994 & 0.762 \\
0.15 & 1.5 & 0.994 & 0.690  & 0.994 & 0.708 \\
0.3 & 0.005 & 0.162 & 0.804  & 0.900 & 0.840 \\
0.3 & 0.15 & 0.872 & 0.778  & 0.986 & 0.814 \\
0.3 & 0.3 & 0.976 & 0.724  & 0.996 & 0.836 \\
0.3 & 0.8 & 0.998 & 0.700  & 0.996 & 0.756 \\
0.3 & 1.5 & 0.986 & 0.680  & 0.988  & 0.680 \\ \hline
\end{tabular}
\end{center}
\end{table}

\begin{table}
\caption{Proportion of the 500 replications where each criterion picked the true model.
Regression coefficient $\beta=0.2$.
Random effects follow a t distribution with 3 degrees of freedom rescaled to have variance $\sigma_u^2$, $u=0,1$.}
\smallskip
\begin{center}
\begin{tabular}{ccrlrc}\hline
 & & \multicolumn{2}{c}{$n_i=80~\forall i$} & \multicolumn{2}{c}{$n_i=320~\forall i$} \\
$\sigma_0^2$ & $\sigma_1^2$ & meanAIC & mAIC & meanAIC & mAIC   \\  \hline
0.005 & 0.005 & 0.122 & 0.786 & 0.838 & 0.822 \\
0.005 & 0.15 & 0.752 & 0.794  & 0.996 & 0.850\\
0.005 & 0.3 & 0.928 & 0.736  & 0.990 &  0.828\\
0.005 & 0.8 & 0.984 & 0.750  & 0.996 & 0.748\\
0.005 & 1.5 & 0.988 & 0.682  & 0.998 & 0.696\\
0.15 & 0.005 & 0.116 & 0.820  & 0.886 & 0.820\\
0.15 & 0.15 & 0.752 & 0.736  & 0.998 & 0.822 \\
0.15 & 0.3 & 0.910 & 0.744  & 0.996 & 0.794\\
0.15 & 0.8 & 0.990 & 0.736  & 0.994 & 0.782 \\
0.15 & 1.5 & 0.990 & 0.716  & 0.992 & 0.762 \\
0.3 & 0.005 & 0.154 & 0.776  & 0.866 & 0.824 \\
0.3 & 0.15 & 0.774 & 0.726  & 0.986 & 0.852 \\
0.3 & 0.3 & 0.944 & 0.748  & 1.000 & 0.806 \\
0.3 & 0.8 & 0.992 & 0.760  & 0.994 & 0.772 \\
0.3 & 1.5 & 0.998 & 0.868  & 0.994 & 0.718 \\ \hline
\end{tabular}
\end{center}
\end{table}

\begin{table}
\caption{Proportion of the 500 replications where each criterion picked the true model.
Regression coefficient $\beta=0.2$.
Random effects follow a normal with mean 0
and variance $\sigma^2_u$, $u=0,1$.}\label{tab:n80beta02normal_random}
\smallskip
\begin{center}
\begin{tabular}{ccrr}\hline
 & & \multicolumn{2}{c}{$n_i\in\{40,80,160\}$} \\
$\sigma_0^2$ & $\sigma_1^2$ & meanAIC & mAIC  \\  \hline
0.005 & 0.005  & 0.156 & 0.790  \\
0.005 & 0.15  & 0.932 & 0.760 \\
0.005 & 0.3   & 0.984 & 0.800  \\
0.005 & 0.8  & 0.994 & 0.796 \\
0.005 & 1.5 & 0.996 & 0.708 \\
0.15 & 0.005 & 0.194 & 0.798 \\
0.15 & 0.15 & 0.926 & 0.746 \\
0.15 & 0.3 & 0.988 & 0.776 \\
0.15 & 0.8 & 0.992 & 0.758 \\
0.15 & 1.5  & 0.990 & 0.752 \\
0.3 & 0.005 & 0.208 & 0.810  \\
0.3 & 0.15  &  0.928 & 0.754 \\
0.3 & 0.3  & 0.988 & 0.774 \\
0.3 & 0.8  & 1.000 & 0.772 \\
0.3 & 1.5  & 0.988 & 0.766 \\ \hline
\end{tabular}
\end{center}
\end{table}

\newpage

\subsection{Covariate description and summary of GLMM fits to the alcohol use data}

Covariates
{\small
\begin{verbatim}
Female: 1 is student is female, 0 if not
Age: age of the student
DoNotLikeSchool: 1 if student does not like school, 0 otherwise
ProudSomeAlot: 1 if student is proud of him/herself a lot or some,
    0 otherwise
Friends: 1 if some of the student's friends have been suspended from
    school, 0 otherwise
FriendsAsk: 1 if student's friend ask student to get drunk some or a
    lot, 0 otherwise
StopYou: 1 if student's family is a lot likely to stop student
    from getting drunk, 0 otherwise
FamilyArgues: 1 if your family argues some or a lot, 0 otherwise
\end{verbatim}
}

Random intercept model
{\small
\begin{verbatim}
Random effects:
 Groups   Name        Variance Std.Dev.
 SchoolID (Intercept) 0.2224   0.4716
Number of obs: 4232, groups:  SchoolID, 25

Fixed effects:
                Estimate Std. Error z value Pr(>|z|)
(Intercept)     -2.54128    0.31952  -7.953 1.81e-15
Female          -0.19667    0.04846  -4.058 4.94e-05
Age              0.10061    0.01941   5.185 2.17e-07
DoNotLikeSchool  0.41888    0.05743   7.294 3.02e-13
ProudSomeAlot   -0.63376    0.05617 -11.284  < 2e-16
Friends          0.42097    0.06091   6.911 4.82e-12
FriendsAsk       1.42403    0.05958  23.901  < 2e-16
StopYou         -0.79153    0.05050 -15.674  < 2e-16
ArguesSomeAlot   0.03555    0.05007   0.710    0.478
\end{verbatim}
}
Model with random intercept and random coefficient
{\small
\begin{verbatim}
Random effects:
 Groups   Name           Variance Std.Dev. Corr
 SchoolID (Intercept)    0.6534   0.8083
          ArguesSomeAlot 0.4340   0.6588   -0.85
Number of obs: 4232, groups:  SchoolID, 25

Fixed effects:
                Estimate Std. Error z value Pr(>|z|)
(Intercept)     -2.57041    0.34870  -7.371 1.69e-13
Female          -0.20338    0.04861  -4.184 2.86e-05
Age              0.09203    0.01954   4.710 2.47e-06
DoNotLikeSchool  0.42672    0.05753   7.417 1.19e-13
ProudSomeAlot   -0.61726    0.05669 -10.889  < 2e-16
Friends          0.40415    0.06095   6.631 3.33e-11
FriendsAsk       1.42107    0.05945  23.904  < 2e-16
StopYou         -0.78243    0.05065 -15.449  < 2e-16
ArguesSomeAlot   0.21190    0.15486   1.368    0.171

\end{verbatim}
}

\end{document}